\documentclass[12pt]{article}
\usepackage{amsmath}
\usepackage{color}
\usepackage{fancyhdr}
\usepackage{verbatim}
\usepackage{amsfonts,longtable,mathrsfs}
\usepackage{epsfig}
\usepackage{amsfonts}
\usepackage{color}
\usepackage[numbers,sort&compress]{natbib}

\def\R{\hbox{{\rm I}\kern-0.2em{\rm R}\kern0.2em}}%mathematical R for reals

\def\bn{\begin{equation}}
\def\en{\end{equation}}
\def\bny{\begin{eqnarray}}
\def\eny{\end{eqnarray}}
\def\be{\begin{eqnarray*}}
\def\ee{\end{eqnarray*}}
\def\bc{\begin{center}}
\def\ec{\end{center}}

\def\({\left(}
\def\){\right  )}
\def\[{\left[}
\def\]{\right]}
\def\bc{\begin{center}}
\def\ec{\end{center}}

\newtheorem{dfn}{Definition}[section]
\newtheorem{thm}{Theorem}[section]
\newtheorem{rem}{Remark}[section]
\newtheorem{pro}{Proposition}[section]

\newtheorem{cor}{Corollary}[section]
\newtheorem{lem}{Lemma}[section]
\newtheorem{exm}{Example}[section]

\def\bn{\begin{equation}}
\def\en{\end{equation}}
\def\bny{\begin{eqnarray}}
\def\eny{\end{eqnarray}}
\def\be{\begin{eqnarray*}}
\def\ee{\end{eqnarray*}}
\def\bdn{\begin{dfn}}
\def\edn{\end{dfn}}
\def\btm{\begin{thm}}
\def\etm{\end{thm}}
\def\bpf{\begin{proof}}
\def\epf{\end{proof}}
\def\bpn{\begin{pro}}
\def\epn{\end{pro}}
\def\brk{\begin{rem}}
\def\erk{\end{rem}}
\def\bcy{\begin{cor}}
\def\ecy{\end{cor}}
\def\blm{\begin{lem}}\def\elm{\end{lem}}
\def\bex{\begin{exm}}
\def\eex{\end{exm}}

 \def\R{{\hat R}}
\begin{document}

\bc {\bf Symmetry, reductions and exact solutions of the difference equation $u_{n+2}=au_n/(1+bu_nu_{n+1})$
  }\ec
\medskip
\bc
Mensah Folly-Gbetoula %\\Mensah.Folly-Gbetoula@wits.ac.za
\footnote{Mensah.Folly-Gbetoula@wits.ac.za} \vspace{1cm}
\\School of Mathematics, University of the Witwatersrand, Johannesburg, South Africa.\\
 %\\e-mails: Mensah.Folly-Gbetoula@wits.ac.za, Abdul.Kara@wits.ac.za \\

\ec
\begin{abstract}
%\vspace{8cm}
\noindent We investigate the solutions of the second-order difference equation %\begin{align}\label{un+2}
$u_{n+2}=(au_n)/(1+bu_nu_{n+1})$
%\end{align}
 using a group of transformations (Lie symmetries) that leaves the solutions invariant.
%We have extended the work of Hydon in \ref{h1}.
\end{abstract}
\section{Introduction} \setcounter{equation}{0}
Symmetry methods for differential equations are well-documented and have been extended to difference equations recently \cite{QR,FK,11,W}. The idea consists of finding symmetries of the equations and use them to lower the order of the equation. Once the solutions of the reduced equations are obtained, one can retrieve the solutions of the original equation by %going up the hierarchy of  change of variables made.\par \noindent
using the invariance of the difference equation under the group of transformations or using the similarity variables.\par \noindent
Some authors have studied the solutions of
\begin{align}\label{xn}
x_{n+1}=\frac{ax_{n-1}}{1+bx_{n-1}x_{n}},
\end{align}
where $a$ and $b$ are real numbers, by putting some restrictions on $a,b$ and the initial conditions $x_{-1}$ and $x_0$. Aloqeili \cite{b-c1/a} investigated the solutions, the stability properties and semi-cycle behavior  of equation (\ref{xn}) when $a=-b=1/A$ with $A\geq 0$, $x_{-1}x_0\neq A^j(1-A)(1-A^j)$, i.e.,
\begin{align}\label{b-c1/a}
x_{n+1}=\frac{x_{n-1}}{A-x_{n-1}x_{n}}.
\end{align}
Cinar \cite{b1c1,b-1c-1,b-1c-b,b1ca,bc} investigated the solutions of
\begin{align}\label{cinar}
\begin{split}
&x_{n+1}=\frac{x_{n-1}}{1+x_{n-1}x_{n}}, \, x_{n+1}=\frac{x_{n-1}}{-1+x_{n-1}x_{n}}, \, x_{n+1}=\frac{x_{n-1}}{-1+Bx_{n-1}x_{n}},\,\\& x_{n+1}=\frac{x_{n-1}}{1+Ax_{n-1}x_{n}}, x_{n+1}=\frac{ax_{n-1}}{1+bx_{n-1}x_{n}}
\end{split}
\end{align}
with the assumptions that
\begin{itemize}
\item $x_{-1}, x_{0}$ are positive real numbers
\item $x_{-1}, x_{0}$ are  real numbers such that $x_0x_{-1}\neq 1$
\item $B\geq 0$, $x_{-1}$ and  $x_{0}$ are  real numbers such that $Bx_0x_{-1}\neq 1$
\item $A,x_{-1}$ and $x_0$ are non-negative real numbers
\item $a,b,x_{-1}$ and $x_0$ are non-negative real numbers,
\end{itemize}
respectively.
\par \noindent
We aim to obtain the solutions of (\ref{xn}) using its symmetry. We expect our solutions to be more general (with less restrictions on $a$ and $b$) and in a \lq single form\rq \, (contrarily to what were presented by these authors).
In order to use our method we will have to \lq shift\rq \, equation (\ref{xn}) and study the equation
\begin{align}\label{un2}
u_{n+2}=\frac{au_{n}}{1+bu_{n}u_{n+1}}
\end{align} instead.
%We expect our solutions to be more general (with less restrictions on $a$ and $b$) and in a 'single form' (contrarily to what were presented by these authors).
There should be  bijections that map our solutions to their solutions. These bijections (if any) will also be investigated.
\subsection{Overview about Lie analysis of difference equations}
Let us consider a $p$th-order difference equation in its general form
\begin{equation}\label{general}
u_{n+p}=\omega(n, u_n, u_{n+1}, \dots, u_{n+p-1}),
\end{equation}
for some function $\omega$, and the point transformations
\begin{equation}\label{Gtransfo}
\Gamma _{\epsilon}: (n,u_n) \mapsto(n,u_n+\epsilon Q (n,u_n)),
\end{equation}
for some continuous function $Q$ which we shall refer to as a characteristic.
\begin{dfn}
The forward shift operator is defined as follows:
\begin{equation}
S:n\mapsto n+1, \qquad S^{i}u_n= u_{n+i}.
\end{equation}
\end{dfn}
The reader can easily check that $\Gamma _{\epsilon}$ is a one-parameter Lie group of transformation admitting
%$\Gamma$ is a one-parameter Lie group of transformations if:
%\begin{itemize}
%\item
%$\Gamma _0$ is the identity map, so that $\tilde{\textbf{x}}=\textbf{x}$ when $\epsilon=0$
%\item
%$\Gamma _{a}\Gamma _{b} = \Gamma _{a+b}$ for every $a$ and $b$ sufficiently close to $0$
%\item
%Each $\tilde{x}_i$ can be represented as a Taylor series in $\epsilon$ (in a neighborhood of $\epsilon =0$ that is determined by $\textbf{x}$), and therefore
%\begin{equation}\label{Gtransfo'}
%\tilde{x}_i(\textbf{x};\epsilon)=x_i+\epsilon \xi_i ({\textbf{x}})+O(\epsilon ^2), i= 1,\dots,p.
%\end{equation}
%\end{itemize}
%In this paper, we shall assume that the Lie point symmetries are of the form
%\begin{equation}\label{Gtransfo''}
%\tilde{n}=n;\quad \tilde{u_n} \simeq  u_n+\epsilon Q(n,u_n)
%\end{equation}
%and that the corresponding infinitesimal generator is given by
\begin{eqnarray}\label{Ngener}
X=  Q\frac{\partial}{ \partial u_n}+ SQ\frac{\partial}{ \partial u_{n+1}} +\cdots +S ^{p-1}Q\frac{\partial}{ \partial u_{n+p-1}}
\end{eqnarray}
as an infinitesimal generator.
\noindent The characteristics $Q=Q(n,u_n,\dots, u_{n+p-1})$ can be found by solving the linearized symmetry condition
\begin{equation}\label{LSC}
\mathcal{S}^{(p)} Q- X \omega=0
\end{equation}
whenever (\ref{general}) holds.
\begin{dfn}
A function $V_n$ is invariant under the Lie group of transformations $\Gamma_{\epsilon}$ if and only if
\begin{equation}
X\left(V_n \right)=0.
\end{equation}
\end{dfn}
Suppose the characteristic $Q$ is known, the invariant $V_n$ can be found by solving the characteristics equation
\begin{equation}
\frac{du_n}{Q}=\frac{du_{n+1}}{SQ}=\dots=\frac{du_{n+p-1}}{S^{n+p-1}Q}\left(=
\frac{dV_n}{0}\right).
\end{equation}
The reader can refer to \cite{11, PO} to deepen his knowledge of how to use symmetry methods for difference equations. %\par \noindent
To the best of our knowledge, there are no packages or computer algebra systems that generate symmetries of difference equations. Often times, the computation becomes cumbersome and some extra ansatz may be needed in order to find the characteristics.
\section{Main results}%{Symmetry and solutions of $u_{n+2}=\frac{au_n}{1+bu_nu_{n+1}}$}
Consider the difference equation (\ref{un2}). Imposing the symmetry condition (\ref{LSC}) we get
\begin{equation}\label{a1}
 Q(n+2,{u_{n+2}})-\frac{a  {u_n}^{2} }{{\left(b {u_n} {u_{n+1}} + 1\right)}^{2}}Q\left(n+1,{u_{n+1}}\right) + {\frac{a }{{\left(b {u_n} {u_{n+1}} + 1\right)}^{2}} } Q\left({n, u_n}\right).
\end{equation}
The latter is an equation containing a function, $Q$, with different arguments making it difficult to solve. To overcome this, we shall assume that $u_{n+1}$ is a function of $n$, $u_n$ and $\omega$. We then proceed by differentiating (\ref{a1}) with respect to $u_n$ (keeping $\omega$  fixed). This leads to %(after clearing the fractions in the resulting equation)
%\begin{flushright}
\begin{align}\label{a2}
\begin{split}
-Q'\left(n+1,{u_{n+1}}\right)+ Q'\left(n,{u_{n}}\right) - \frac{2}{u_n}Q\left(n,{u_{n}}\right)=0.
\end{split}
\end{align}
%\end{flushright}
By differentiating (\ref{a2}) with respect to $u_{n}$ (keeping $u_{n+1}$ fixed) we obtain
\begin{align}\label{a5}
\begin{split}
 Q''\left(n,{u_{n}}\right) - \frac{2}{u_n}Q'\left(n,{u_{n}}\right) + \frac{2 }{{u_n}^{2}}Q\left(n,{u_{n}}\right)=0
\end{split}
\end{align}
whose solution is given by
\begin{align}\label{a6}
\begin{split}
Q\left(n,{u_{n}}\right) = \alpha (n) {u_n}^2 +\beta (n) {u_n}
\end{split}
\end{align}
for some functions $\alpha$ and $\beta$ of $n$.\par The last step will consist of substituting (\ref{a6}) in (\ref{a1}) to get the symmetry given by
\begin{eqnarray}\label{gener}
X=  (-1)^n u_n \partial u_n- (-1)^n u_{n+1} \partial_{u_{ n+1}}.
\end{eqnarray}
It is easy to check that the function
\begin{equation}\label{vnun}
v_n=u_nu_{n+1}
\end{equation}
is invariant under $X$ given in (\ref{gener}) and that
\begin{equation}\label{vn+1}
v_{n+1}=\frac{av_n}{1+bv_n}.%\qquad \text{with}\quad v_0=u_0u_1\neq -1/c.
\end{equation}
The solution of (\ref{vn+1}) is given by
\begin{equation}\label{vn}
v_n=
\begin{cases}%\label{solvn}
 \frac{a-1}{c_1 \left(\frac{1}{a}\right)^{n-2}-c_1 \left(\frac{1}{a}\right)^{n-1}-b \left(\frac{1}{a}\right)^n+b} \qquad \text{if}\qquad  a\neq 1,\\ \\
\frac{1}{bn+c_0} \qquad \text{if} \qquad a=1
\end{cases}
\end{equation}
for some constants $c_0$ and $c_1$.
%%%%%%%%%%%%%%%%%%%%%%%%%%%%%%
\subsection{Case $a=1$}
When $a=1$ equation (\ref{un2}) becomes
\begin{equation}\label{b1}
u_{n+2}=\frac{u_n}{1+bu_nu_{n+1}}.
\end{equation}
In this case we are saying that the solution of (\ref{vn+1}) is given by
\begin{equation}\label{solvnb1}
v_{n}=\frac{1}{bn+c_0}.
\end{equation}
Invoking (\ref{vn}), we have that
\begin{equation}\label{solvnb1'}
u_{n+1}=\frac{1}{(bn+c_0)u_n}.
\end{equation}
\textbf{Note}: \textit{The order of equation (\ref{b1}) has been reduced by one.} \newline
The solution of (\ref{solvnb1'}) given by
\begin{align}
u_n =& \exp \left((-1)^{n-1} c_2+(-1)^{n-1} \sum_{k_1 =0}^{n-1}(-1)^{-k_1} \ln \left|v_{k_1}\right|\right)\\
 =&\exp \left((-1)^{n-1} c_2+(-1)^{n-1} \sum_{k_1 =0}^{n-1}-(-1)^{-k_1} \ln \left|c_0+b k_1\right|\right),
\end{align}
where $c_2$ is an arbitrary constant, is also the solution of (\ref{b1}). It has to be noted that $c_0=\frac{1}{u_0u_1}$ and $c_2=\ln \left|\frac{1}{u_0}\right|$ in this case. Therefore, the most general solution of (\ref{b1}) is given by
\begin{subequations}
\begin{align}%\label{sol=1}
& u_n = \exp \left((-1)^{n-1} \ln \left|\frac{1}{u_0}\right|+(-1)^{n-1} \sum_{k_1 =0}^{n-1}(-1)^{-k_1} \ln \left|\frac{u_0u_1}{1+bu_0u_1k_1}\right|\right),\label{unb11}\\
& \quad = \exp \left((-1)^{n-1} \ln \left|{u_1}\right|+(-1)^{n-1} \sum_{k_1 =1}^{n-1}(-1)^{-k_1} \ln \left|\frac{u_0u_1}{1+bu_0u_1k_1}\right|\right),\label{unb13}\\
&\text{with} \quad  -1/(bu_0u_1) \notin \{1,2,\dots n-1\}.\label{unb12}
\end{align}
\end{subequations}
%with $-1/(bu_0u_1) \notin \{0,1,2,\dots n-1\}$
\begin{rem}
The solution (\ref{unb11}) can be split into.
\begin{equation}
u_n=
\begin{cases}%\label{solvn}
 u_0\frac{\prod_{t=0}^{n/2}[1+2tbu_0u_1 ]}{\prod_{t=0}^{n/2}[1+(2t+1)bu_0u_1 ]} \qquad \text{n even}\\ \\
 u_1\frac{\prod_{t=0}^{\frac{n-1}{2}-1}[1+(2t+1)bu_0u_1 ]}{\prod_{t=0}^{\frac{n-1}{2}}[1+2tbu_0u_1 ]} \qquad \text{n odd}
\end{cases}
\end{equation}
\begin{itemize}
\item
If we let $x_n =u_{n+1}, u_0=k, u_1=h, a=1 $ and  $b=A$, we get the result
\begin{equation}
x_n=
\begin{cases}%\label{solvn}
 k\frac{\prod_{i=0}^{\frac{n+1}{2}-1}[2Ahki +1]}{\prod_{i=0}^{\frac{n+1}{2}-1}[(2i+1)Ahk+1 ]} \qquad \text{n odd}\\ \\
 h\frac{\prod_{i=0}^{\frac{n}{2}-1}[1+(2i+1)Ahk ]}{\prod_{i=0}^{\frac{n}{2}}[2iAhk+1 ]} \qquad \text{n even}
\end{cases}
\end{equation}
 obtained by C. Cinar in \cite{b1ca} for
 \begin{align}
 x_{n+1}= \frac{x_{n-1}}{1+Ax_nx_{n-1}}
 \end{align}
 and his restriction ($x_{-1},x_0$ and $A$ are positive real numbers) is a special case of our restriction given in (\ref{unb12}), that is,$ -1/(Ax_{-1}x_0) \notin \{1,2,\dots n-1\}$ in this case.
%\end{rem}
\item
If we let $x_n =u_{n+1}, u_0=k, u_1=h, a=1, b=1$, we get the result
\begin{equation}
x_n=
\begin{cases}%\label{solvn}
 k\frac{\prod_{i=0}^{\frac{n+1}{2}-1}[2hki +1]}{\prod_{i=0}^{\frac{n+1}{2}-1}[(2i+1)hk+1 ]} \qquad \text{n odd}\\ \\
 h\frac{\prod_{i=0}^{\frac{n}{2}-1}[1+(2i+1)hk ]}{\prod_{i=0}^{\frac{n}{2}}[2ihk+1 ]} \qquad \text{n even}
\end{cases}
\end{equation}
 obtained by C. Cinar in \cite{b1c1} for
\begin{align}
 x_{n+1}= \frac{x_{n-1}}{1+x_nx_{n-1}}
 \end{align}
 and his restriction ($x_{-1}$ and $x_0$ are positive real numbers) is a special case of our restriction given in (\ref{unb12}), that is,$ -1/(x_{-1}x_0) \notin \{1,2,\dots n-1\}$ in this case.
\end{itemize}
\end{rem}
\subsection{ Case $a\neq 1$}
We mentioned earlier that the solution of (\ref{vn+1}) when $a\neq1$ is given by
\begin{align}\label{vnbc1}
v_n=& \frac{a-1}{c_1 \left(\frac{1}{a}\right)^{n-2}-c_1 \left(\frac{1}{a}\right)^{n-1}-b \left(\frac{1}{a}\right)^n+b}.%\\
%=& \frac{(b-1)u_0u_1b^n}{b-1+cu_0u_1(b^n-1)}\\
%=&\frac{u_0u_1b^n}{1+cu_0u_1(\sum_{i=0}^{n-1}b^i)}
\end{align}
Here, $c_1=1/au_0u_1$ and the above equation simplifies to
\begin{align}\label{vnb}
v_n=& \frac{(a-1)u_0u_1a^n}{a-1+bu_0u_1(a^n-1)}\nonumber\\
=&\frac{u_0u_1a^n}{1+bu_0u_1(\sum_{i=0}^{n-1}a^i)}.
\end{align}
Thank to (\ref{vnun}) we have $u_{n+1}=v_n/u_n$ and then,
%\begin{subequations}
\begin{align}\label{unb}
u_n=&\exp{\left[ (-1)^{n-1}\ln\left|\frac{1}{u_0}\right|+(-1)^{n-1}\sum_{k_1=0}^{n-1}(-1)^{-k_1} \ln|v_{k_1}|   \right]},\nonumber\\
%=&\exp{\left[ (-1)^{n-1}\ln\left|\frac{1}{u_0}\right|+(-1)^{n-1}\sum_{k_1=0}^{n-1}(-1)^{-k_1} \ln\left|\frac{u_0u_1a^{k_1}}{1+bu_0u_1(\sum_{i=0}^{k_1-1}a^i)}\right|   \right]},\nonumber\\
=&\exp{\left[ (-1)^{n-1}\left(\ln\left|{u_1}\right|+\sum_{k_1=1}^{n-1}(-1)^{-k_1} \ln\left|\frac{u_0u_1a^{k_1}}{1+bu_0u_1(\sum_{i=0}^{k_1-1}a^i)}\right|  \right) \right]}
\end{align}
%u_n=&\exp \Bigg((-1)^{n-1} \Bigg[ \ln\left(\frac{1}{u_0}\right)+\nonumber \\&\sum_{k_1=0}^{n-1}(-1)^{-k_1} \log\left(\frac{b-1}{\frac{1}{bu_0u_1} \left(\frac{1}{b}\right)^{k_1-2}-\frac{1}{bu_0u_1} \left(\frac{1}{b}\right)^{k_1-1}-c \left(\frac{1}{b}\right)^{k_1}+c}\right)   \Bigg]\Bigg)
\begin{align}\label{unb'}
%\text{ where }  v_n  \text{ is given by } (\ref{vnb})
 \text{ with} -1/(bu_0u_1) \notin \{1,1+a,\dots, \sum_{i=0}^{n-2} a^{i}\}. \qquad \qquad \qquad \qquad \qquad \quad
\end{align}
%\end{subequations}
\begin{rem}
The solution (\ref{unb}) can be split into
\begin{equation}
 u_n=
\begin{cases}%\label{solvn}
% u_0\frac{\prod_{t=0}^{n/2-1}v_{2t+1}}{\prod_{t=0}^{n/2-1}v_{2t}}=
u_0a^{n/2}\frac{\prod_{t=0}^{n/2-1}{\left((a-1)+bu_0u_1(a^{2t}-1)\right)}}{\prod_{t=0}^{n/2
 -1}{\left((a-1)+bu_0u_1(a^{2t+1}-1)\right)}}= u_0a^{n/2}\frac{\prod_{t=1}^{n/2-1}{\left(1+bu_0u_1\sum_{i=0}^{2t-1}a^i\right)}}{\prod_{t=0}^{n/2
 -1}{\left(1+bu_0u_1\sum_{i=0}^{2t}a^i\right)}}\;\text{n even}\\ \\
%{u_1}\frac{\prod_{t=1}^{\frac{n-1}{2}}v_{2t}}{\prod_{t=0}^{\frac{n-1}{2}-1}v_{2t+1}} =

u_1a^{\frac{n-1}{2}}\frac{\prod_{t=0}^{\frac{n-1}{2}-1}{\left((a-1)+bu_0u_1(a^{2t+1}-1)\right)}}{\prod_{t=1}^{\frac{n-
1}{2}}{\left((a-1)
 +bu_0u_1(a^{2t}-1)\right)}}
=u_1a^{\frac{n-1}{2}}\frac{\prod_{t=0}^{\frac{n-1}{2}-1}{\left(1+bu_0u_1\sum_{i=0}^{2t}a^i\right)}}{\prod_{t=1}^{\frac{n-1}{2}}{\left(1
 +bu_0u_1\sum_{i=0}^{2t-1}a^i\right)}} \; \text{n odd}
\end{cases}
\end{equation}
%%%%%%%%%%%%%%%%%%%%%%%%%%%%%%%%%%%
\begin{itemize}
\item
If we let $x_n =u_{n+1}, u_0=k, u_1=h$, $a=-1$ and  $b=-1$, we get the result
\begin{equation}
%x_n=
\begin{cases}%\label{solvn}
 x_{2t+1}=\frac{k}{(hk-1)^{t+1}}\\ \\
 x_{2t+2}=h(hk-1)^{t+1}
\end{cases}
\end{equation}

 obtained by C. Cinar in \cite{b-1c-1} for
 \begin{align}
 x_{n+1}= \frac{x_{n-1}}{-1+x_nx_{n-1}}
 \end{align}
and his restriction ($x_{-1}x_0 \neq 1$ ) coincides with our restriction given in (\ref{unb'}), that is, $ 1/(x_{-1}x_0) \notin \{1\}$ in this case.
%\end{rem}
\item
If we let $x_n =u_{n+1}, u_0=k, u_1=h, a=-1$ and $b=-B$, we get the result
\begin{equation}
%x_n=
\begin{cases}%\label{solvn}
 x_{2t+1}=\frac{k}{(Bhk-1)^{t+1}}\\ \\
 x_{2t+2}=h(Bhk-1)^{t+1}
\end{cases}
\end{equation}
 obtained by C. Cinar in \cite{b-1c-b} for
\begin{align}
 x_{n+1}= \frac{x_{n-1}}{-1+Bx_nx_{n-1}}
 \end{align}
and his restriction ($Bx_{-1}x_0 \neq 1$ ) coincides with our restriction given in (\ref{unb'}), that is, $ 1/(Bx_{-1}x_0) \notin \{1\}$ in this case.
%%%%%%%%%%%%%%%%%%%%%%%%%%%%%%%%%to check
  \item
If we let $x_n =u_{n+1}, a=-b=1/A$, we get the result
\begin{equation}
x_n=
\begin{cases}%\label{solvn}
 x_0\prod_{i=1}^{\frac{n}{2}}\frac{A^{2i-1}(1-A)-(1-A^{2i-1})x_{-1}x_0}
 {A^{2i}(1-A)-(1-A^{2i})x_{-1}x_0} \qquad \text{n even}\\ \\
 x_{-1}\prod_{i=0}^{\frac{n+1}{2}-1}\frac{A^{2i}(1-A)-(1-A^{2i})x_{-1}x_0}
 {A^{2i+1}(1-A)-(1-A^{2i+1})x_{-1}x_0} \qquad \text{n odd}
\end{cases}
\end{equation}
 obtained by  Aloqeili in \cite{b-c1/a} for
\begin{align}
 x_{n+1}= \frac{x_{n-1}}{A-x_nx_{n-1}}
 \end{align}
  and his restrictions ($A\geq 0$, $x_{-1}x_0 \neq A^j(1-A)(1-A^j)$ ) is a special case of our restriction given in (\ref{unb'}), that is, $ A/(x_{-1}x_0) \notin \{1,1+A^{-1},\dots, \sum_{i=0}^{n-2} A^{-i}\}$ in this case.
%\end{itemize}
 \item
If we let $x_n =u_{n+1}, u_0=k, u_1=h$, we get the result
\begin{equation}
%x_n=
\begin{cases}%\label{solvn}
 x_{2t+1}= ka^{t+1}\frac{\prod_{i=0}^{t-1}{\left(1+bu_0u_1\sum_{j=0}^{2i+1}a^j\right)}}{\prod_{i=0}^{t
 }{\left(1+bu_0u_1\sum_{j=0}^{2i}a^j\right)}}\\ \\
 x_{2t+2}= ha^{t+1}\frac{\prod_{i=0}^{t}{\left(1+bu_0u_1\sum_{j=0}^{2i}a^j\right)}}{\prod_{i=0}^{t
 }{\left(1+bu_0u_1\sum_{j=0}^{2i+1}a^j\right)}}
\end{cases}
\end{equation}
 obtained by C.Cinar in \cite{bc} for
\begin{align}
 x_{n+1}= \frac{ax_{n-1}}{1+bx_nx_{n-1}}
 \end{align}
 and his restriction ($a,b,  x_{-1}$ and $x_0$ are non-negative real numbers ) is a special case of our restriction given in (\ref{unb'}), that is, $ -1/(bx_{-1}x_0) \notin \{1,1+a,\dots, \sum_{i=0}^{n-2} a^{i}\}$ in this case.
\end{itemize}
\end{rem}
\section{Conclusion}
%\section{Conclusion}
We have used  symmetry methods for difference equations to solve equation (\ref{un2}). The solutions were given in (\ref{unb}), that is,
\begin{equation*}
u_n=\exp{\left[ (-1)^{n-1}\ln\left|{u_1}\right|+(-1)^{n-1}\sum_{k_1=1}^{n-1}(-1)^{-k_1} \ln\left|\frac{u_0u_1a^{k_1}}{1+bu_0u_1(\sum_{i=0}^{k_1-1}a^i)}\right|   \right]}
\end{equation*}
with $-1/(bu_0u_1) \notin \{1,1+a,\dots, \sum_{i=0}^{n-2} a^{i}\}$. %\par \noindent
Contrary to the results obtained in papers \cite{b-c1/a,bc}, our solutions are  \lq single\rq \,solutions (with less restrictions on $a$ and $b$). For the sake of clarification, we made some change of variables and some assumptions to show that their solutions are obtained by splitting our solutions into two categories depending on the parity of $n$.

\end{document}